\documentclass[preprint,authoryear]{elsarticle} 
\usepackage{graphicx}
\usepackage{hyperref}
\usepackage{caption} 

\usepackage{amssymb}
\usepackage{amsmath}
\usepackage{amsthm}
\newtheorem{thm}{Theorem}
\newdefinition{define}{Definition}

\begin{document}

\title{The Realization of 3D Topological Spaces Branched Over Graphs}

\author[1]{Christopher L Duston}

\affiliation[1]{organization={Department of Mathematics and Physics, Merrimack College},
addressline={315 Turnpike St},
postcode={01845},
city={N Andover},
country={USA}}
\ead{dustonc@merrimack.edu}


\begin{abstract}
  In this paper we present an implementation of a computer algorithm that automatically determines the topological structure of spacetime, using a branched covering space representation. This algorithm is applied to a few simple examples in dimension 3, and a complete set of the topological spaces branched over several graphs are found. We also include some new visualizations of the branched covering construction, in order to aid and clarify the understanding of how these structures can be used in quantum gravity to realize the topological nature of the spacetime foam.
  
\end{abstract}


\maketitle
\section{Introduction}\label{s:Intro}
The role that topology plays in theoretical physics, General Relativity (GR) in particular, is an unclear one. On one hand, there is the foundational work of \citet{Hawking-Ellis-1973} on the large-scale and causal structure of spacetime, and corresponding observations coming from the cosmic microwave background \citep{Levin-etal-1998}. On the other hand, we have several no-go theorems regarding the topology change of spatial sections of a Lorentzian manifold \citep[e.g.][]{Anderson-DeWitt-1986}. Observationally, although exotic options have been explored \citep{Luminet-etal-2003, Duston-2017}, the general consensus is that the topology of the Universe is something akin to $\mathbb{S}^3\times \mathbb{R}$.

However, the story is quite different in quantum gravity. The essential philosophical principle is that just as quantum fields vary wildly on small scales, the same should be true of the geometrical structure of the Universe, and that these wild fluctuation should extend to topology as well (often described as a \textit{spacetime foam}). Further, a defining feature of GR is \textit{background independence}; the theory does not exist on a fixed background, but rather the background is determined by the dynamics of the theory. However, this background independence does not apply to topology as GR is only a local theory - it does not provide us with a mechanism to directly probe the topological structure. The topology must be set ahead of time by the theorist, a situation which is also true of the dimension as well as the differentiable structure (an issue that is more well studied in recent years, see \citet{Asselmeyer-Maluga-Brans-2007,Asselmeyer-Maluga-Brans-2014, Duston-2011, Duston-2021}). 

This expectation about the behavior of quantum gravity needs to be compared to the reality of specific modern approaches. Prime among them is Loop Quantum Gravity (LQG), which is well-developed enough for introductory textbooks to be dedicated to it \citep[e.g.][]{Thiemann-2007}. Another interesting approach is Causal Dynamical Triangulations (CDT), which is a direct attempt to simulate the thermal physics of discrete spacetimes \citep{Ambjorn-Durhuus-Jonsson-1997,AGJL-2012}. In both of these approaches, however, the spatial topology is fixed to be trivial, and is not allowed to change - thus, the full vision of the spacetime foam described in the previous paragraph is explicitly not realized. Even in the case of Causal Set Theory, which aims to build up spacetime from fundamentally topological units, it is not clear that the foam structure is manifest \citep{Surya-2008}.

With this background in hand, we now discuss a new approach that proposes to include changing topologies into quantum gravity, LQG and CDT in particular. It is based on representing the spatial sections (or the spacetime manifold as a whole) as branched covering spaces. This approach has been studied in LQG \citep[\textit{topspin networks},][]{DMA, Duston-2012, Duston-2020}, and a few explicit calculations have been done \citep{Duston-2013-arXiv,Villani-2021}. We are working on moving these ideas to CDT, but the statistical tools used in that approach pose different challenges then the analytical calculations that can be successfully completed in LQG. This paper will address this technical problem by demonstrating an implementation of an algorithm that can ``track'' the topology in the critical dimension 3, and will apply these calculations over a particular collection of graphs. The choice of the graphs are for demonstration purposes, and therefore somewhat arbitrary, but we will present all the topologies that can be realized over some of these graphs at a fixed order of the cover. This result is a technical one, but absolutely required in order to make progress including topology change in quantum gravity via branched covering spaces.

The rest of this paper is organized as follows: \S\ref{s:Background} will present the key mathematical construction we will be using, as well as introduce several new graphical clarifications. \S\ref{s:Model} will present the specifics of the model, and some of the technical details about the software used. \S\ref{s:Results} will present the results of the simulations, and we will finish in \S\ref{s:Conclusion} by summarizing and giving some perspective on where we envision this project going from here.

\section{Spacetime as a Branched Covering Spaces}\label{s:Background}
Most of the pertinent technical details on this material can be found in \citet{DMA} or \citet{Duston-2012}, so here we will just sketch the main ideas, and try to add some new clarity to the conceptual understanding of these structures. The key classical result upon which the construction is based is the following \citep{Alexander-1920}:

\begin{thm}[Alexander's Theorem]\label{thm:Alexander}
Any compact oriented 3-manifold can be described as a branched covering of $\mathbb{S}^3$, branched along a graph.
\end{thm}

We can roughly describe this result as presenting a way to reparameterize the geometric and topological degrees of freedom for arbitrary manifolds in such a way that we can be guaranteed to describe them all. It can be generalized to dimension 4 \citep{Piergallini-1995}, the number of covers can be restricted \citep{Hilden-1976}, and one can find complete sets branched over special types of knots \citep{PS}. A key feature of these listed results that should be noted here is that the branch locus is codimension 2 - smaller choices of the codimension will typically not result in the covers being well-defined manifolds.

\begin{figure}
  \begin{center}
    \includegraphics[scale=0.3]{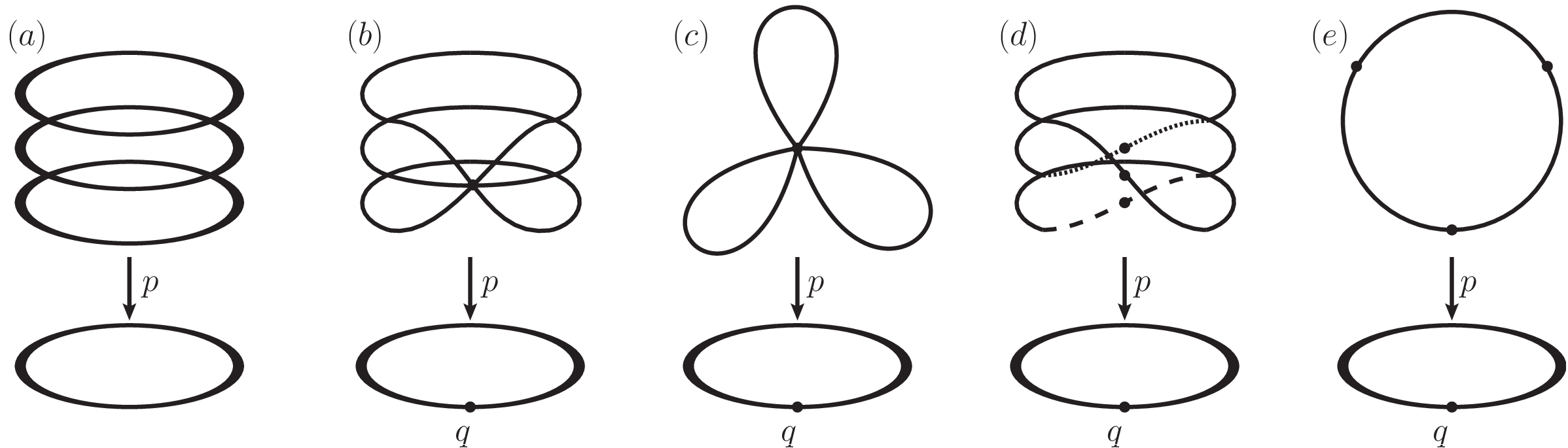}
    \caption{Several examples of order 3 branched covering spaces over $\mathbb{S}^1$. (a) shows an unbranched cover, $\mathbb{S}^1\bigcup\mathbb{S}^1\bigcup\mathbb{S}^1$. (b) shows a branched cover with valence 1, and (c) illustrates how that space is not a manifold. (d) shows an unbranched cyclic cover, (e) illustrates how this last space is simply $\mathbb{S}^1$ in the ``cover-view''.}\label{fig:Dim1Ex}
  \end{center}
  \end{figure}

The first of our conceptual sketches can be found in Figure \ref{fig:Dim1Ex}, which shows both unbranched and branched 1-dimensional 3-fold covers over a sphere $\mathbb{S}^1$. Note that in the unbranched case, the number of inverse images $p(U)$ of an open set $U\in\mathbb{S}^1$ is constant (that is, three), but in the first branched example this fails over a particular point $q$ (the ``branch locus'', and $p^{-1}(q)$ is the ``ramification locus''). We've chosen this illustration to also demonstrate that in this codimension (one), you do not generally get nice manifolds, and also that the manner in which these constructions are presented can obscure their true identity (in (d) and (e) the cyclic cover $\mathbb{S}\to\mathbb{S}$ is actually an unbranched manifold).

\begin{figure}
  \begin{center}
    \includegraphics[scale=0.35]{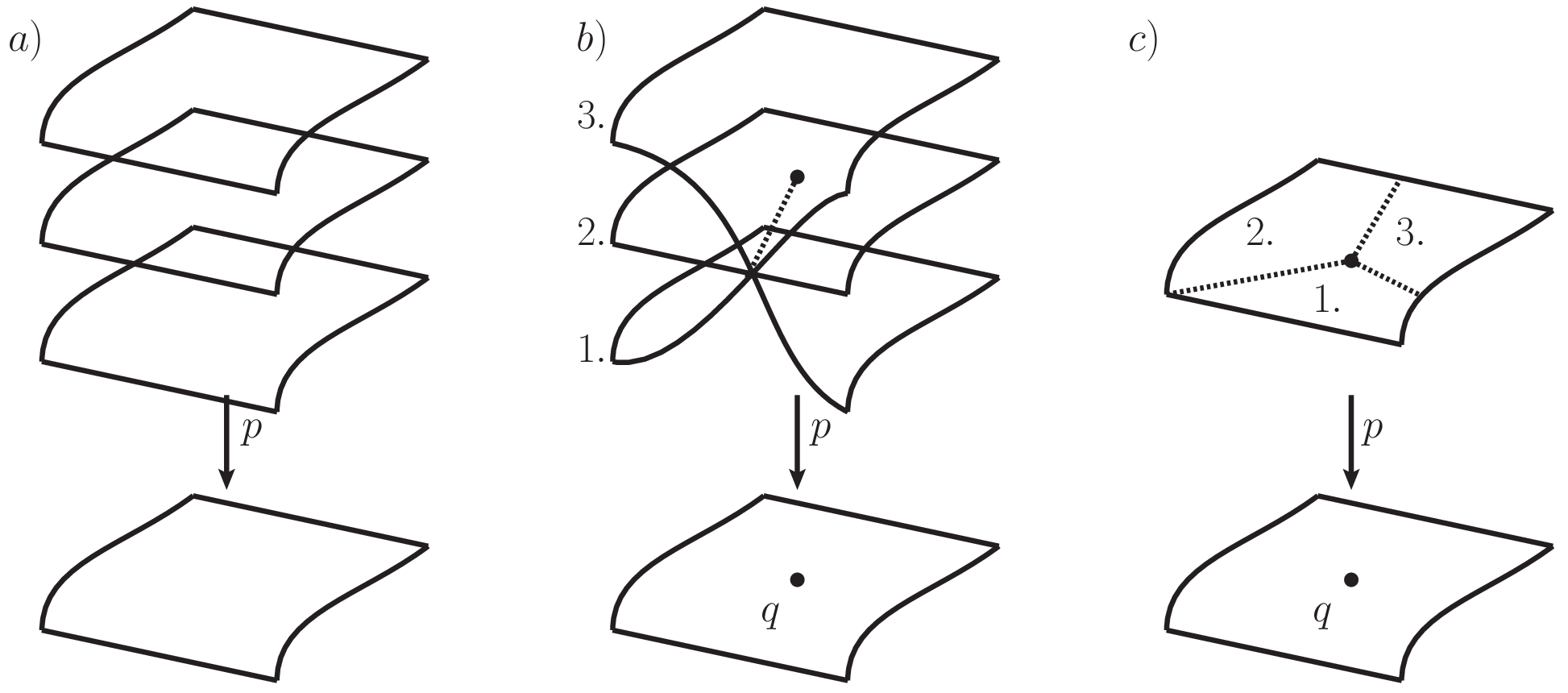}
    \caption{Several examples of order 3 branched covering spaces over $\mathbb{S}^2$. (a) shows an unbranched cover, $\mathbb{S}^2\bigcup\mathbb{S}^2\bigcup\mathbb{S}^2$. (b) shows a branched cover with valence 1. We freely acknowledge that in this case, the illustration does not fully capture how the sheets ``are sewn together''. We address this directly in the next figure, and (c) illustrates how this branched cover is simply a copy of $\mathbb{S}^2$.}\label{fig:Dim2Ex}
  \end{center}
  \end{figure}

By moving to dimension two, we can illustrate the codimension two situation (Figure \ref{fig:Dim2Ex}). In this case, the ``base-view'' in (b) looks very much like a non-manifold structure, but the ``cover view'' in (c) shows that the branching is actually contained in the map $p$, not the cover itself\footnote{We do not know if there are conventional labels for these representations, and are just proposing the ``base view'' and ``cover view'' offer some additional clarity.}. In dimension 2, prototypical examples of these objects are the Riemann surfaces, typically described as $n$th roots of the complex plane,

\[S=\{(w,z)\in\mathbb{C}^2|w^n=z\}.\]
In this way we see the problem with our Figure \ref{fig:Dim2Ex}(b); the surfaces cannot be embedded in $\mathbb{R}^3$, but are represented as graphs in $\mathbb{C}^2$. For an excellent introduction to this perspective on Riemann surfaces, we suggest \citet{Teleman-2003}.

\begin{figure}
  \begin{center}
    \includegraphics[scale=0.25]{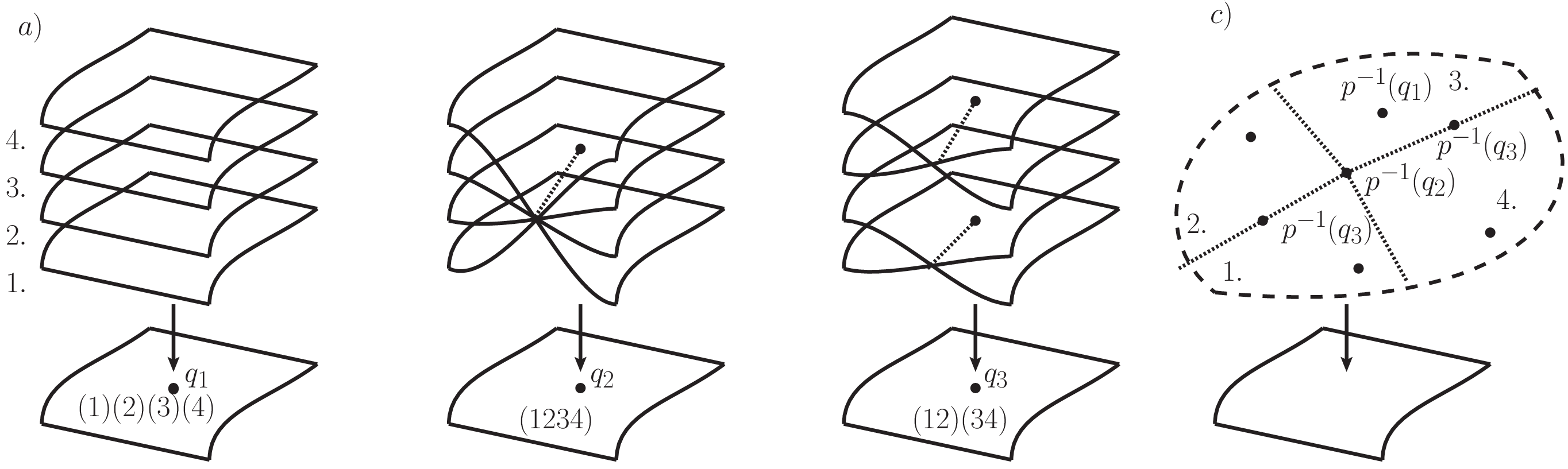}
    \caption{Several examples of regions of order 4 branched covering spaces over $\mathbb{S}^2$. (a) shows three regions around branch loci, of valency 4, 1, and 2, respectively. Figure (c) provides a view of a region of this cover including all the inverse images of the branch points. 
    }\label{fig:Dim2Ex_2}
  \end{center}
  \end{figure}

Of course, the branch locus does not have to be restricted to a single point; in Figure \ref{fig:Dim2Ex_2} we show three regions of such a space with a single branch point in each, and also irregular valency $v_q$ (number of solutions) over the branch point. Again, we demonstrate in \ref{fig:Dim2Ex_2}(c) that this seemingly complicated surface is just $\mathbb{S}^2\to\mathbb{S}^2$, and the branching is a property of the map, not the space.

One classifies such examples with a branch index $b_q$, which enumerates the ``missing solutions'' in the inverse image of the covering map, and is related to the valency via $b_q=n-v_q$ for an $n$-sheeted cover. In dimension 2, the total branch index $b$ tells us all the topological information about an $n$-fold cover $S$ over a base $B$ via the Riemann-Hurwitz formula,

\[\chi(S)=n\chi(B)-b.\]
This formula is trivial to see in this dimension, as $\chi=V-F+F$ for a chosen triangulation. 

At this point it should be emphasized what is being implied by the branched cover construction, which is explicit in Alexander's Theorem - the covering space can be completely reconstructed with knowledge of the base (an $n$-sphere, in our case) and knowledge of the local representation of the covering map. There is a convenient way to parameterize this structure, using a labeling of the vertices with elements of the permutation group $\mathcal{S}_n$ (we believe this parameterization of the \textit{monodromy} is due to \citet{Piergallini-1993}), and we have included those in Figure \ref{fig:Dim2Ex_2}. The intuition behind these labels is immediate - as a path travels around the branch point, crossing the branch cut, the path transitions from one sheet to another. This transition is described by an action of a permutation (element of $\mathcal{S}_n$), as shown in Figure \ref{fig:Dim2Ex_2}. There are conditions on the choice of these permutation elements, based on global consistency, which we discuss later.

\begin{figure}
  \begin{center}
    \includegraphics[scale=0.35]{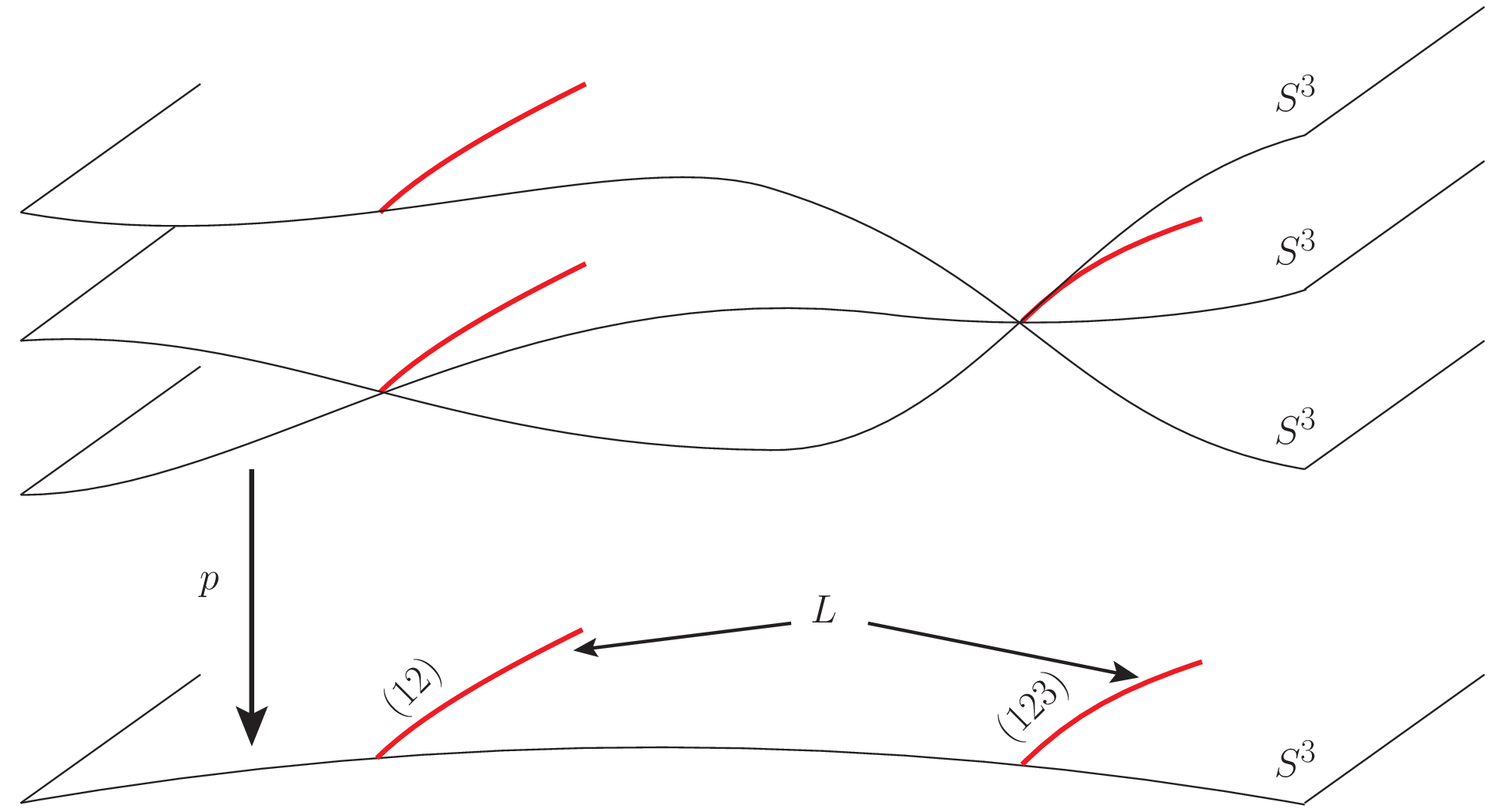}
    \caption{A schematic of a 3-fold branched cover over a 3-sphere. This figure was adopted from \citet{Duston-2013}.}\label{fig:Dim3Ex}
  \end{center}
  \end{figure}

Finally, we discuss the dimension 3 case in Figure \ref{fig:Dim3Ex}, which will occupy the rest of this work. Note that while we can't represent the $\mathbb{S}^3$ sheets well on the page, we've tried to make it clear that the branch locus is still codimension 2; a 1-dimensional submanifold. These are also labeled by elements of the permutation group, with the interpretation that traveling in a path around the locus is what causes the transitions between sheets. We hope the significance of the earlier figures in lower dimensions are clear at this point - although the cover appears to be very pathological, all the branching is occurring in the map, not in the covering space itself. Roughly speaking, in codimension 2 there is ``enough room'' to travel around ramification locus in the cover, transversing the individual sheets and returning to the initial point without encountering any non-manifold behavior. The covering spaces are parts of spheres, glued along codimension 2 submanifolds.

\section{Statistical Model}\label{s:Model}
With the technical details regarding branched covering spaces in hand, we move on to the statistical model we are presenting. It is motivated by a particular approach to quantum gravity, CDT, as discussed in \S\ref{s:Intro}. However, our main goal is to illustrate the concrete realization of these spaces. Therefore we will not be allowing the branch locus to be changing, and will only be altering the topological labels. Specifically, we are going to choose relatively simple graphs as the branch loci, and establish what kinds of topologies we can detect over them. We do point out that as we discussed earlier, such restrictions to a finite number of covers (even 3, see \citet{Hilden-1976}) may not restrict the available topological structures at all.

\begin{figure}
\begin{minipage}{0.32\textwidth}
    \begin{center}
      \includegraphics[scale=0.30]{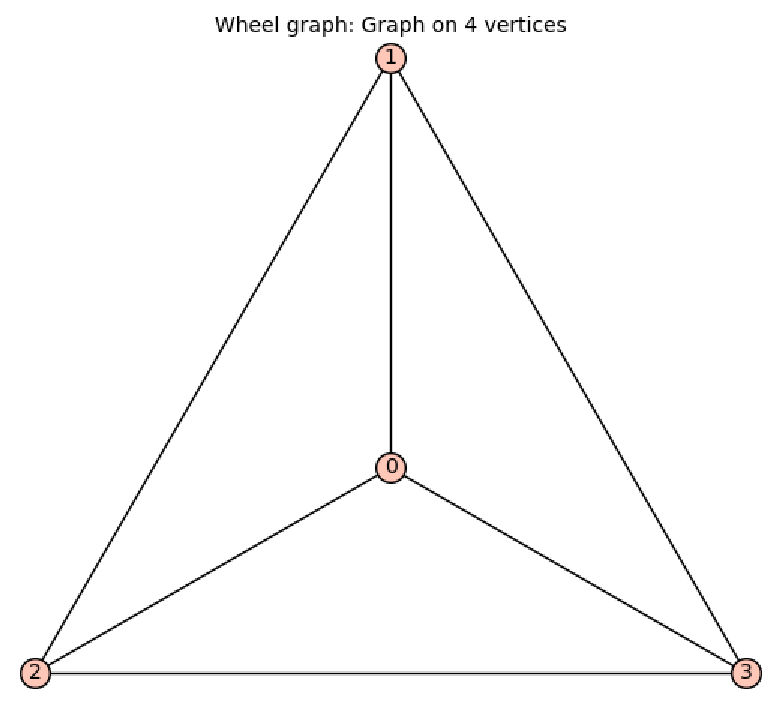}
    \end{center}
  \end{minipage}
  \begin{minipage}{0.32\textwidth}
    \begin{center}
      \includegraphics[scale=0.30]{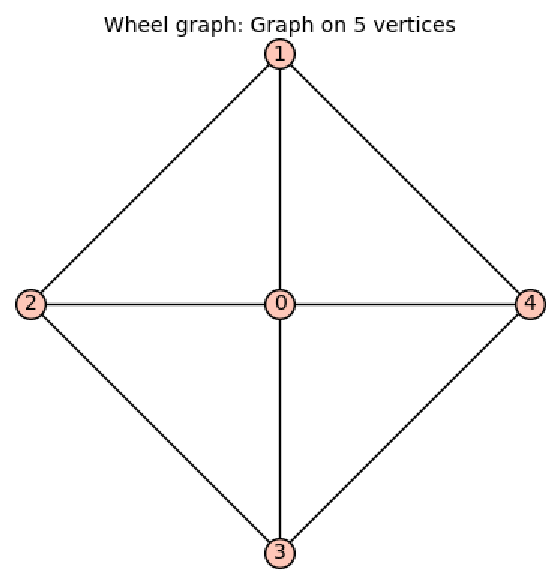}
    \end{center}
  \end{minipage}
  \begin{minipage}{0.32\textwidth}
    \begin{center}
      \includegraphics[scale=0.30]{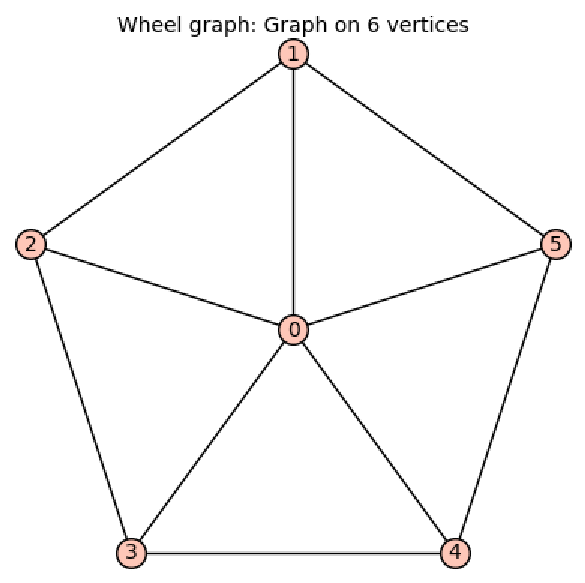}
    \end{center}
  \end{minipage}
  \caption{The first three wheel graphs $W_n$.}\label{fig:WheelGraphs}

  \end{figure}

Our primary examples are going to be the wheel graphs $W_n$, and are constituted as a single vertex with edges connected to every other vertex of an $n-1$ cycle, examples of which are shown in figure \ref{fig:WheelGraphs}. Although the first wheel graph ($W_4$) does match the tetrad models often used in quantum gravity (LQG specifically, see \citet{Villani-2021}), the choice is rather arbitrary. They do represent simple, understandable graphs, and we do not expect this specific choice generally impacts our results in what follows.

In CDT, the space of quantum states is typically explored with a Monte Carlo-style algorithm, by stepping around elements of phase space and recording parameters of the system. We have the advantage that the total number of topologies in our system is finite and relatively small, specifically $|\mathcal{S}_n|^E$, since each of the edges can be labeled with an element of $\mathcal{S}_n$. Of course, not all of these arrangements of edges will be valid, but this is the maximum number. Thus, we can simply check them all.

  The conditions on the permutations are the Wirtinger relations - in our context, where there are intersections but no crossings, this is simply done by picking a direction around vertices (say, clockwise) and ensuring the product of all of the edges is unity, $\sigma_1^{\pm 1}\sigma_2^{\pm 1}\sigma^{\pm 1}_3...=1$, where the $+1$ is used to ingoing edges and the $-1$ for outgoing edges. Our simulations are going to find all of the possible topologies that can be realized over particular graphs, and in the next section we will discuss the technical tools used to describe and analyze this system.

  \section{Technology: SageMath and GAP}
  The primary technology used in this work is SageMath \citep{sagemath}, in particular functions related to graph theory that allow the user to reference, modify, and analyze graphs. We would like to present in more detail how we have used the implementation of the Groups, Algorithms, and Programming \citep{GAP} software package in SageMath, because it gives us the ability to determine some physical parameters of these spaces which we would not otherwise have easy access to.

  Our routine defines the graph $\Gamma$ and creates an initial free group $H$, with the same number of elements as edges in the graph. We then create the ``graph group''\footnote{This terminology is apparently not standard, but means the same thing as ``knot group''.} $G$ - the fundamental group of the complement $\pi_1(\mathbb{S}^3\setminus \Gamma)$ - using the Wirtinger relations in the free group (for details around this and other aspects of topology and knots, we recommend the excellent text of \citet{Fox-1962}). For $W_4$, this is equivalent to

  \[G=H/[a_5a_1^{-1}a_4=1,~a_3a_2a_1=1,~a_6^{-1}a_3^{-1}a_5^{-1}=1,~a_6a_4^{-1}a_2^{-1}=1],\]
for $a_i$ generators in $H$. GAP can simplify this set - in particular, it can determine that these 4 relations are not independent, and actually only sets 3 of the generators in $H$.

  Now to create all the covers, we assign permutation labels to all the edges. These must respect the same Wirtinger relations as the elements of the free group, so we can cycle through all possible ways of choosing 3 of them, and then use the relations to set the other 3. For these examples, we will be using a 3-fold cover, so the number of possible topologies branched over $W_n$ will be $3!^{n-1}$.

  To determine the actual topologies realized, we will utilize an algorithm first presented in \citet{Fox-1962}, and discussed in more detail in \citet{Duston-2013-arXiv}. The essential idea is to add generators to the fundamental group in the cover $\pi_1(p^{-1}(\mathbb{S}^3\setminus \Gamma))$ that models how the sheets of the cover are ``glued together'', and then add generators corresponding to moving along the graph $p^{-1}(\Gamma)$. This gives us the fundamental group of the cover. To our knowledge, our implementation of this algorithm is the first one done on a modern computer.

However, these are now presented as finite groups - a list of generators and a list of relations. Since some of these might be different presentations of the same group, we'd like to compare them to each other. This is an example a ``word problem'' in group theory, and is generally an undecidable problem. However, GAP can at least tell us how big each group is, in terms of number of generators. In addition, we can ask it to \textit{try} and solve the word problem using the routine \verb|StructureDescription|, and we find that for these relatively small groups, it does a reasonably good job\footnote{We should be very clear that the GAP manual says this function ``is not intended to be a research tool, but rather an educational tool'', so we present it here merely as demonstration - more advanced techniques will need to be developed for these classifications to be considered more robust.}.

\section{Results of Simulations}\label{s:Results}

\begin{figure}
\begin{minipage}{0.49\textwidth}
    \begin{center}
      \includegraphics[width=\textwidth]{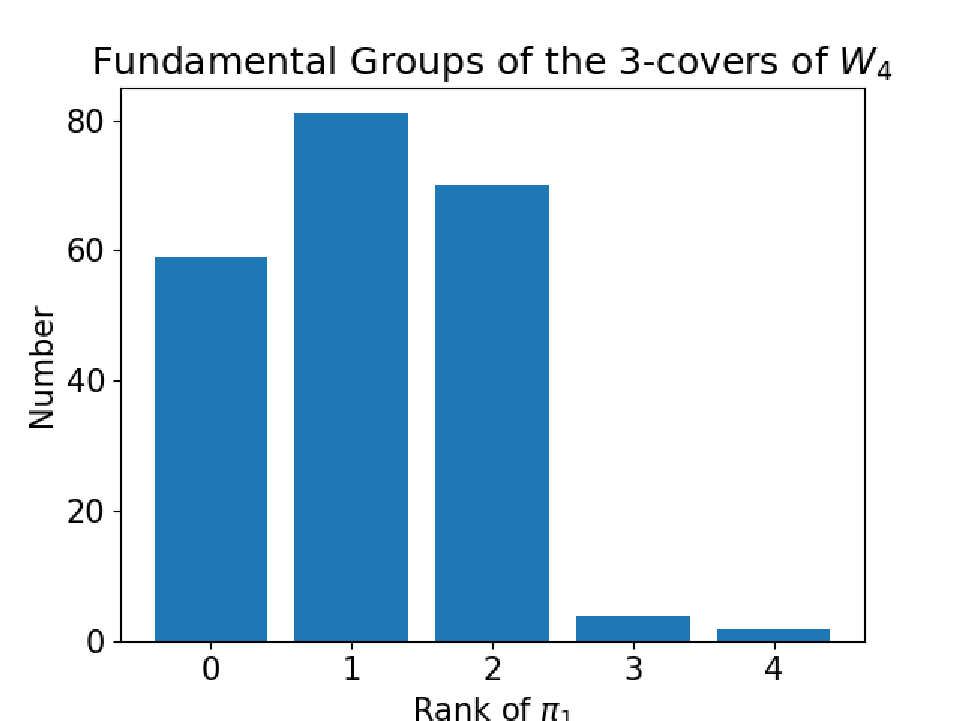}
    \end{center}
  \end{minipage}
  \begin{minipage}{0.49\textwidth}
    \begin{center}
      \includegraphics[width=\textwidth]{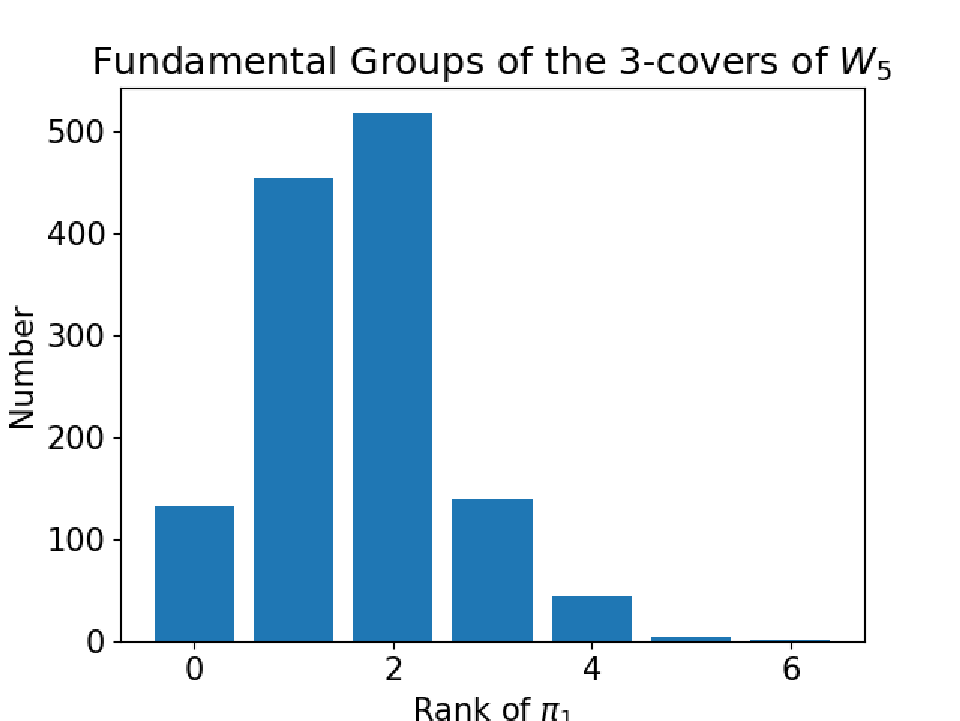}
    \end{center}
  \end{minipage}
  \caption{The complete set of fundamental groups, distributed by rank, that can be realized as 3-fold branched coverings over the wheel groups $W_4$ and $W_5$.}\label{fig:Wheel-1}
  \end{figure}
  
\begin{figure}
\begin{minipage}{0.49\textwidth}
    \begin{center}
      \includegraphics[width=\textwidth]{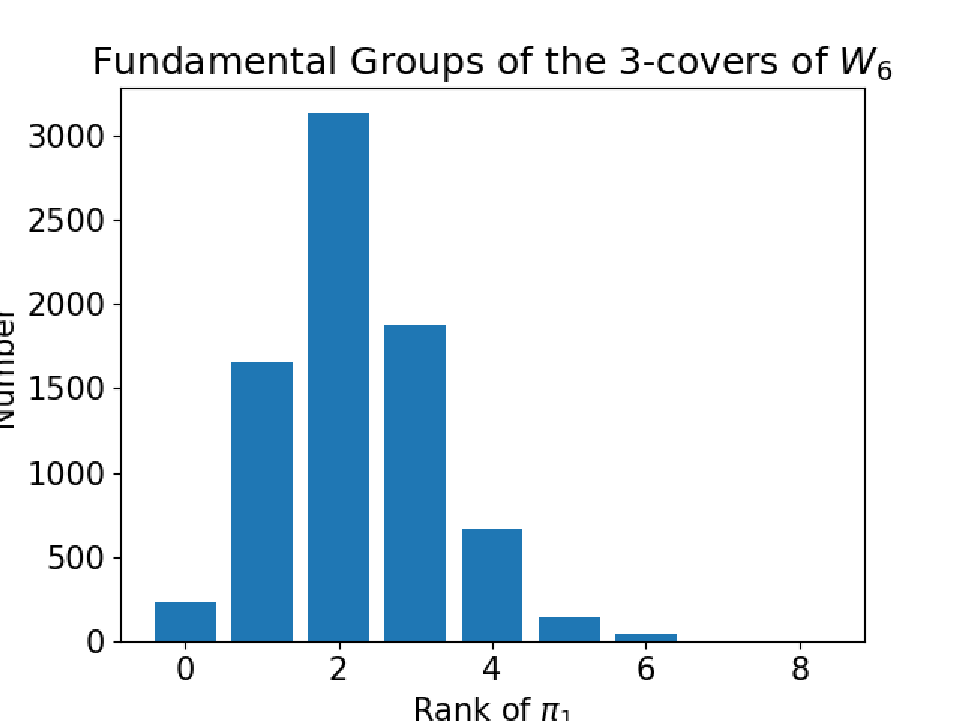}
    \end{center}
  \end{minipage}
  \begin{minipage}{0.49\textwidth}
    \begin{center}
      \includegraphics[width=\textwidth]{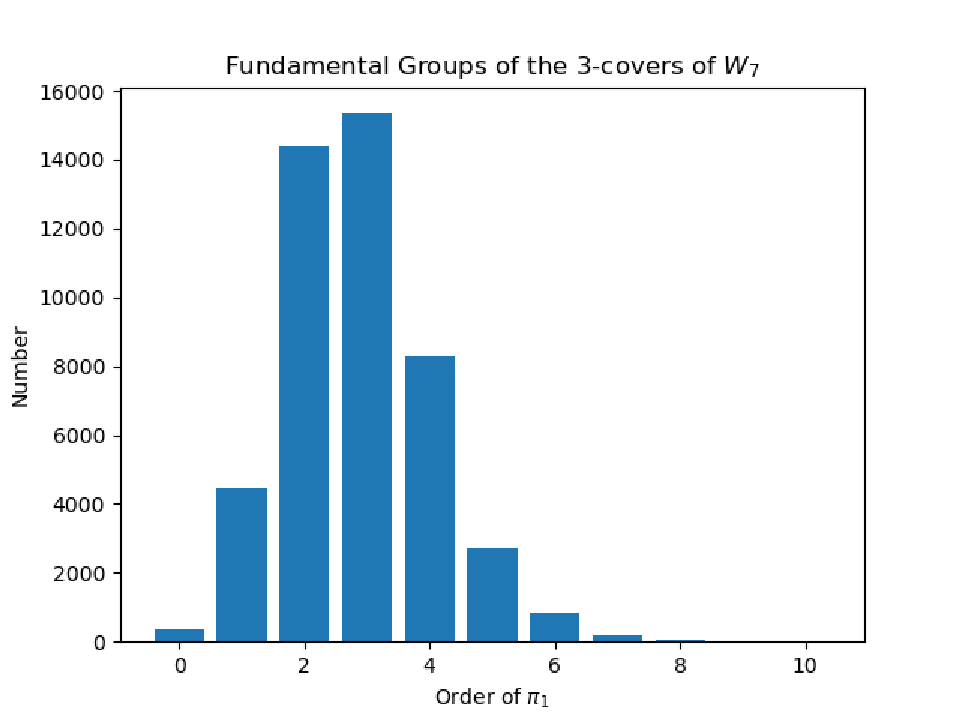}
    \end{center}
  \end{minipage}
   \caption{The complete set of fundamental groups, distributed by rank, that can be realized as 3-fold branched coverings over the wheel groups $W_6$ and $W_7$.}\label{fig:Wheel-2}
\end{figure}

\begin{figure}\begin{center}
    \includegraphics[width=\textwidth]{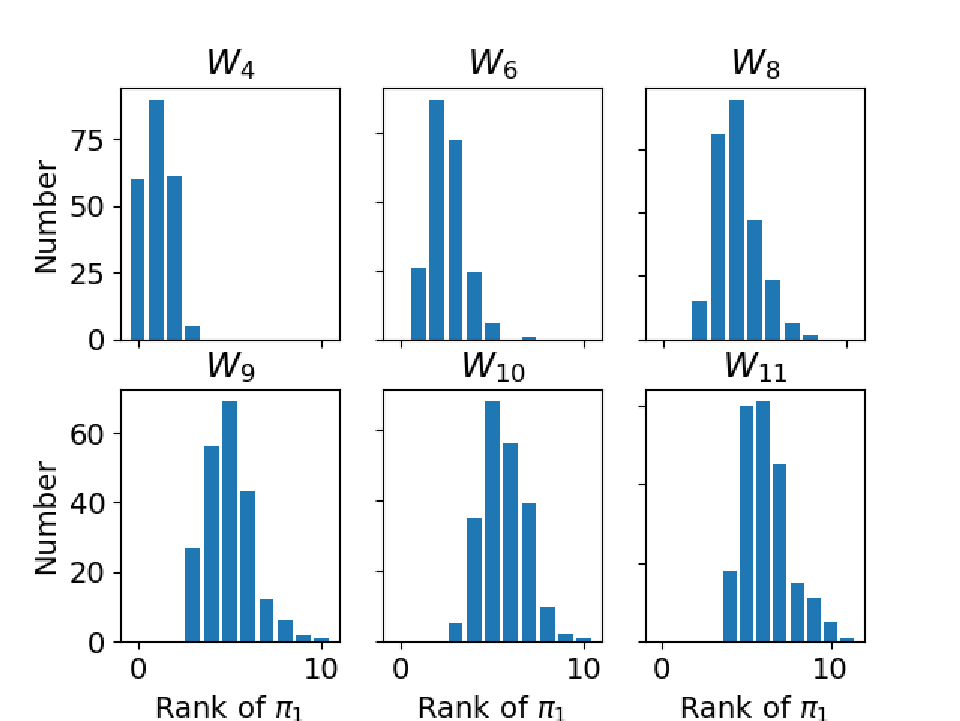}
    \caption{This figure fixes the number of topologies found to be 216, to illustrate the dependence of the size of the fundamental group over more complicated graphs.}\label{fig:Wheel-3}
   \end{center}\end{figure}

The first results of this study, the fundamental groups that can be realized as 3-fold covers of the first few wheel graphs, are shown in Figures \ref{fig:Wheel-1} and \ref{fig:Wheel-2}. This behavior matches our rough expectations - there are far more simple fundamental groups than complex ones, and the smaller ones are harder to produce when the complexity of the graph increases. These first figures show all the fundamental groups that can be created over $W_n$ for $n=4,5,6,7$, but to illustrate the distribution for higher $n$ we include Figure \ref{fig:Wheel-3} as well.

\begin{figure}\begin{center}
\begin{minipage}{0.32\textwidth}
    \begin{center}
      \includegraphics[width=\textwidth]{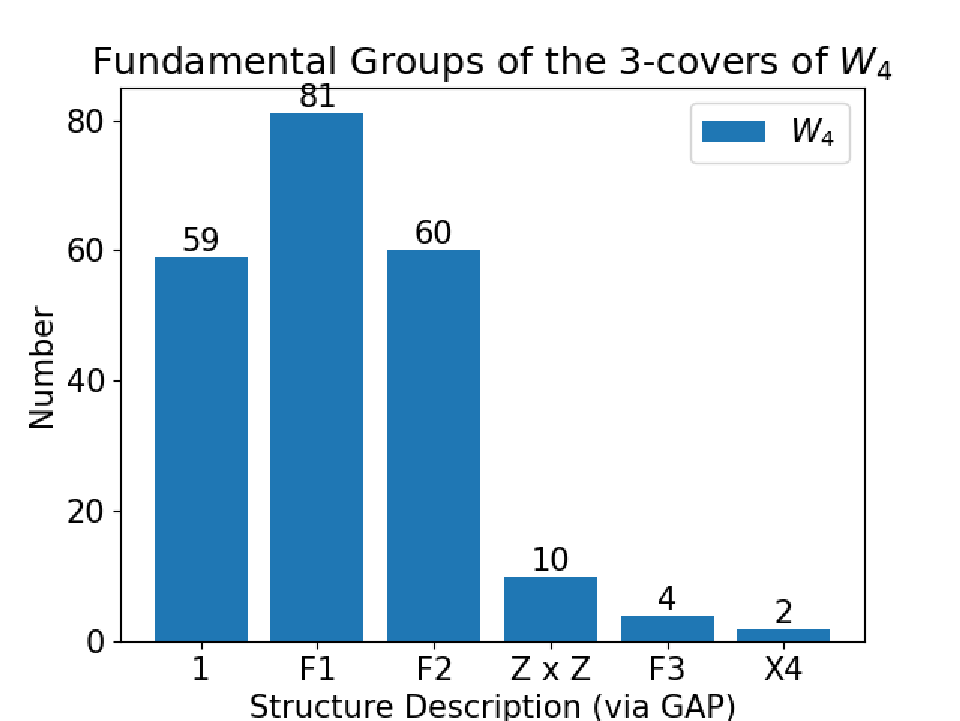}
    \end{center}
  \end{minipage}
  \begin{minipage}{0.32\textwidth}
    \begin{center}
      \includegraphics[width=\textwidth]{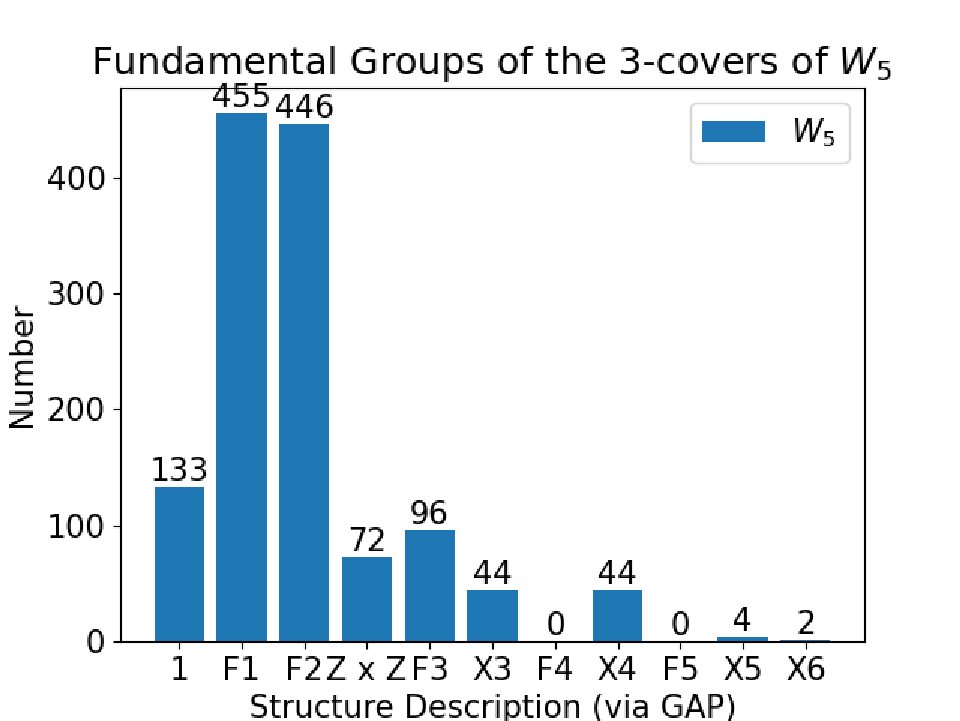}
    \end{center}
  \end{minipage}
  \begin{minipage}{0.32\textwidth}
    \begin{center}
      \includegraphics[width=\textwidth]{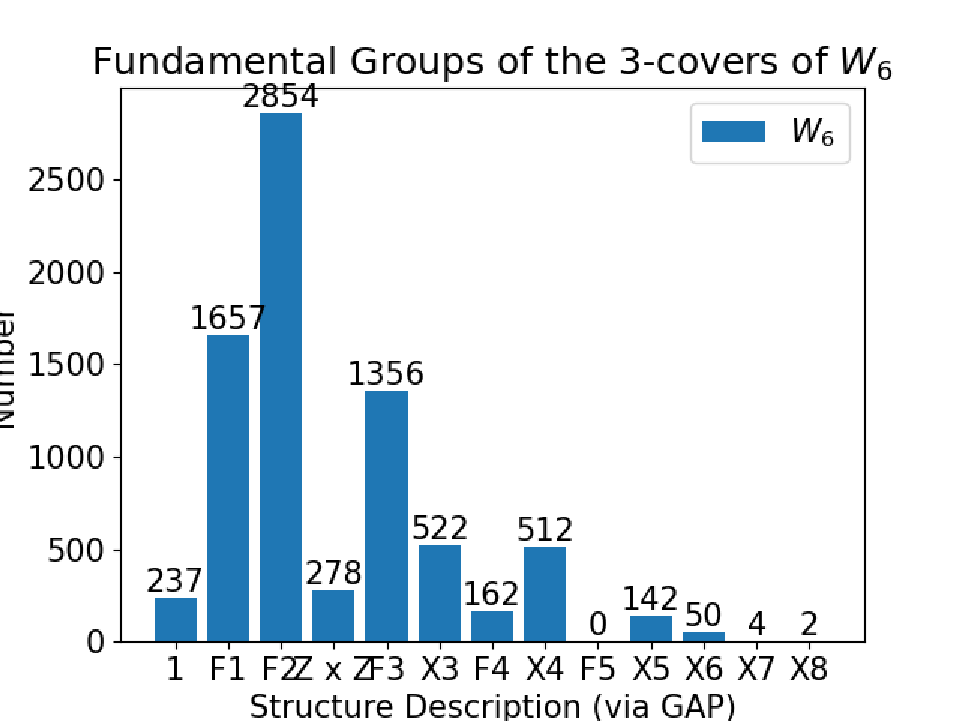}
    \end{center}
  \end{minipage}
  \caption{This figure illustrates the ability of GAP to differentiate between fundamental groups of different ranks. $Fn$ is the free group on $n$ elements, whereas $X_n$ is an unknown group on $n$ elements. The ``degeneracy lifting'' occurs for example, in $F_2$ vs $\mathbb{Z}\times \mathbb{Z}$ over all three graphs.}\label{fig:WheelSplit}
  \end{center}\end{figure}

The 3-manifolds in Figure \ref{fig:Wheel-1} and \ref{fig:Wheel-2} are differentiated by the rank of their fundamental groups. This is a good start, but far from a complete classification because the rank is only a partial invariant (groups with different ranks are not isomorphic, but equal rank does not imply isomorphism). As discussed in the previous section, we can use GAP to partially lift this degeneracy, which we demonstrate in Figure \ref{fig:WheelSplit}. A further improvement would be to use the second homology group $H_2=\pi_1/[\pi_1,\pi_1]$ to further differentiate between fundamental groups with similar rank (for a nice introduction to the topology of 3-manifolds, we recommend \citet{Hatcher-2007}). We will have to be satisfied with this, as the complete classification of 3-manifolds is still an open problem \citep{Thurston-2014}.

\begin{figure}
\begin{minipage}{0.49\textwidth}
    \begin{center}
      \includegraphics[width=\textwidth]{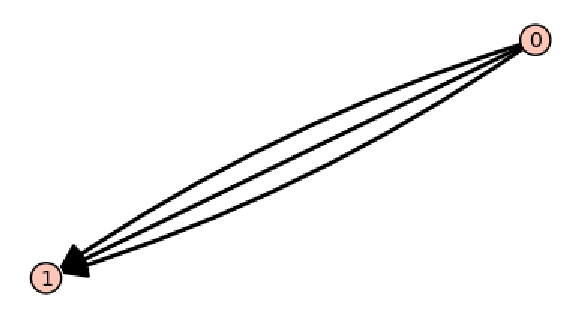}
    \end{center}
  \end{minipage}
  \begin{minipage}{0.49\textwidth}
    \begin{center}
      \includegraphics[width=\textwidth]{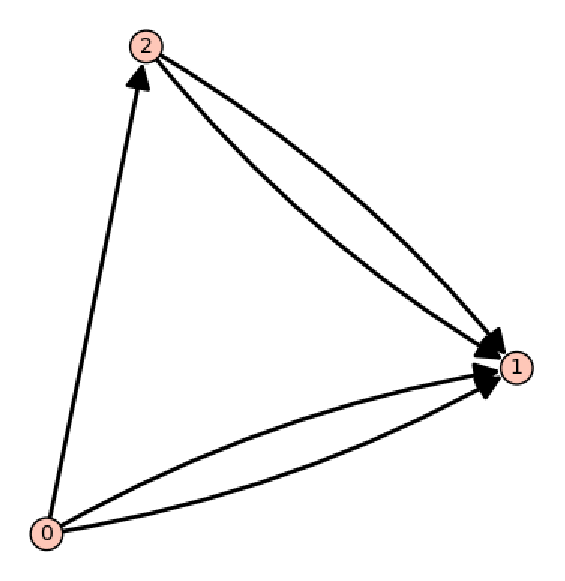}
    \end{center}
  \end{minipage}
   \caption{}\label{fig:BubbleEx}
\end{figure}

For the purposes of illustration, we've applied the algorithm to another family of graphs, shown in Figure \ref{fig:BubbleEx}. The primary difference is that these are multigraphs, with multiple edges between two vertices. The inspiration here is that these graphs are related to those used in \citet{Rovelli-Vidotto-2008} to model weakly inhomogeneous cosmology. Unlike the the wheel graphs, it is not a formally defined family of graphs, independent of this inspiration. 

\begin{figure}
\begin{minipage}{0.49\textwidth}
    \begin{center}
      \includegraphics[width=\textwidth]{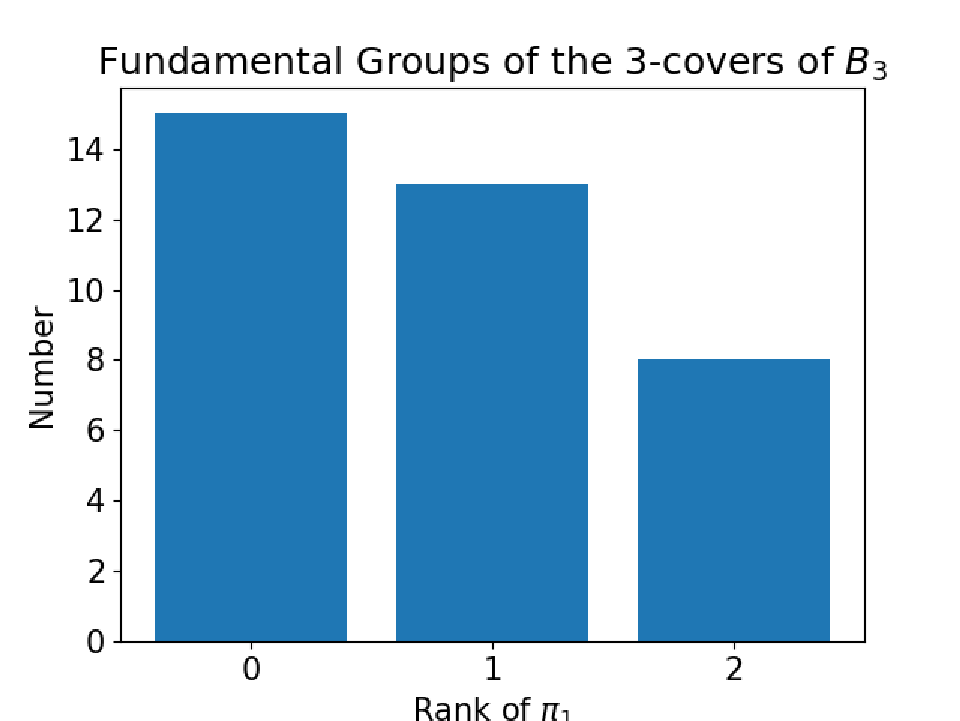}
    \end{center}
  \end{minipage}
  \begin{minipage}{0.49\textwidth}
    \begin{center}
      \includegraphics[width=\textwidth]{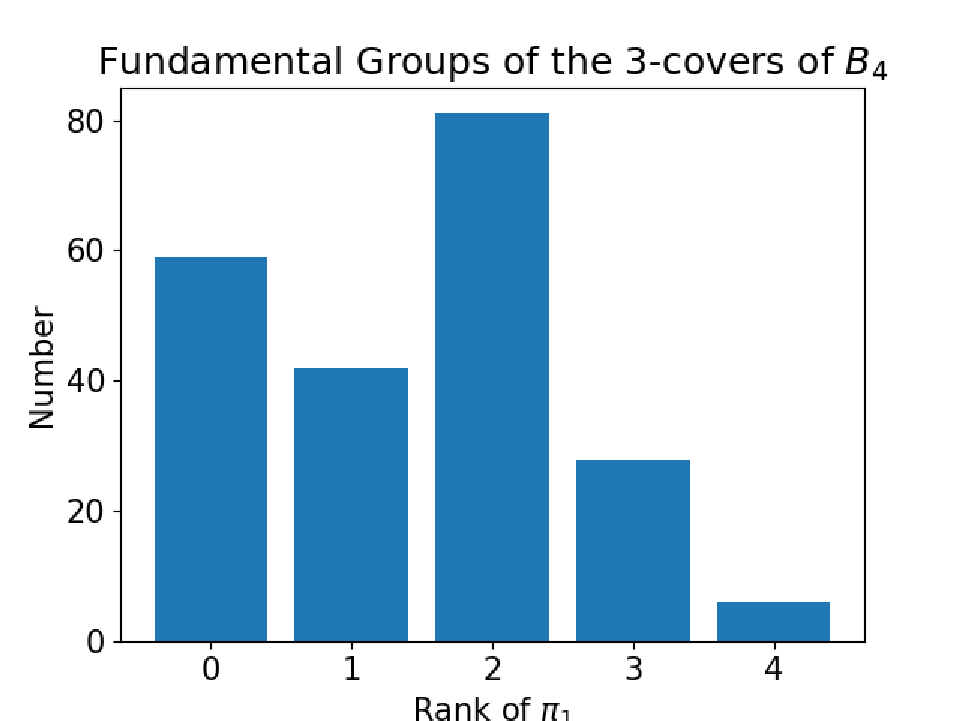}
    \end{center}
  \end{minipage}
   \caption{}\label{fig:Bubble-2}
\end{figure}

The complete set of topological spaces that can be realized over 3-fold coverings of these is shown in Figure \ref{fig:Bubble-2}. We don't consider the differences between these and the wheel graphs as substantive - the realization of a particular spacetime over a particular graph is probably mostly related to the number of vertices and edges present in the graph, not any special properties of the graphs.

\section{Summary and Outlook}\label{s:Conclusion}
The main goal of this paper was to demonstrate a key technical requirement of the statistical approach to studying topology in quantum gravity, the algorithmic determination of topological structure. This general requirement exists for any approach to the ``spacetime foam'' which includes topology, although our specific solution only applies to the branched covering space reparametrization. It was not at all obvious to us that it was possible to realize something like the Fox Algorithm on a computer system, which would be required to automatically determine an appropriate topological description of the spacetime. It turns out, because of the existence of tools like SageMath and (in particular) GAP, it was relatively straightforward to implement directly, \textit{i.e.} in the symbolic language.

We then used this routine to find all the branched covers that could be realized over some special graphs as branch loci, the wheel graphs $W_n$.  As comparison, we did this for another family of graphs, inspired by models of weakly inhomogeneous quantum cosmology. We do not claim that understanding the topological structures over these particular graphs is significant on their own, just that it was not possible to answer such questions before the computer implementation existed. This statement is emphasized by Figure \ref{fig:WheelSplit}, in which the structure of some of the fundamental groups are identified by GAP.

Our next goal in this area is to directly apply this routine to the problem of topology in CDT. With a (partial) specification of the topology, it is now possible to start asking (model-dependent) questions like ``what was the topology of the early Universe?'', ``could `small' topological structures exist in the Universe today?'', or ``was there a mechanism that drove the Universe to be topologically trivial?'' in the context of CDT. This same routine could be used in LQG, following the by-hand approaches from \citet{Duston-2013-arXiv} and \citet{Villani-2021}.

Of course, one could also use this routine to study branched covering spaces over graphs and knots themselves. For example, \citet{PS} have proven that the Borromeo rings represent a \textit{Universal Link}, in that any compact, oriented 3-manifold without boundary can be realized as a branched covering space over them. With our routine, one could imagine statistically looking for more such Universal Links, or study general properties that such links (or graphs, knots, \textit{etc.}) are suspected to have. Generalizing to higher dimensions would be relatively straightforward (as long as one kept the codimension of the locus to be 2, as required by Alexander's theorem), so extensions of \textit{e.g.} Piergallini's results in dimension 4 \citep{Piergallini-1995}, or of higher dimensional knots more generally \citep{Ogasa-2018}, are statistically accessible using this routine.

\bibliography{mybib}

\end{document}